Lecture Notes of the Autumn School on
Correlated Electrons 2023

Eva Pavarini and Erik Koch (Eds.)

# Orbital Physics in Correlated Matter

Autumn School organized by
the Institute for Advanced Simulation
at Forschungszentrum Jülich
18 – 22 September 2023

# 9 Strong Correlations at Oxide Interfaces What is Hidden in a Plane View?

Jak Chakhalian

Rutgers University

136 Frelinghuysen Rd, Piscataway, NJ 08854

## Contents

A prime goal for this lecture is to provide you with a reasonably self-sufficient answer to the question of what interesting effects can happen if you join two dissimilar materials with correlated carriers to construct a 'sandwich' with the interface across those layers. Through the lens of physical phenomena, we will delve into the design ideas that lead to the creation of new synthetic quantum materials with properties primarily governed by the interface.

About the structure of the lecture: After introducing key concepts from the physics of correlated electrons, I switch to the guiding notions for building new synthetic materials with properties unattainable in bulk. Next, I briefly discuss the nucleation and growth of thin films based on the pulsed laser deposition method (PLD) or laser molecular beam epitaxy (MBE). After that, I illustrate those design ideas by several recent examples, ranging from a correlated polar metal to a quantum spin liquid. The lecture concludes with a list of ten currently *unsolved problems* that are worth further exploration.

# 1   Primer on the physics of correlated oxides.

**Why transition metal oxides?** Transition metal ions (TM) are commonly found in complex oxides, which make up the largest group of crystals on Earth. Besides oxygen, these compounds contain an element from the $d$-series in the periodic table, specifically $3d$, $4d$, or $5d$ TM oxides. In contemporary notation, many complex oxides with TM ions belong to the family of *quantum materials with correlated electrons* [1]. In general, it is the variation in the outermost $d$-shell configuration of these elements that gives rise to the great complexity in the crystal structures, electronic properties, and magnetic interactions in TMOs. There are many informative reviews on this topic, but as a one-stop source I recommend [2] as a comprehensive resource.

**What crystal structures exist, and why are they formed?** TMOs have numerous types of crystal structures, spanning all seven crystal systems! Although the most rigorous language of determining a crystal structure is to identify a conventional unit cell and the corresponding space group, going through all 230 space groups is certainly not the purpose of this lecture. Instead, as TMOs are predominately regarded as *ionic* crystals, the driving force of stabilizing a specific structure is the lattice energy: namely, for an ion located in a lattice, it experiences an overall electrostatic potential from the other ions (both cations and anions), which is the so-called Madelung potential $V_I$. The associated electrostatic energy of the ion is the product of its net charges with the Madelung potential. By summing over all ionic sites, we calculate the lattice energy. Broadly speaking, to determine the minimal value of the lattice energy we need to know the specific details of the crystal structure. However, since the Coulomb interaction between cation and anion is attractive, it is natural to assume that the lattice energy dramatically decreases as more anions surround a cation. On the other hand, as we place more and more anions near the cation, the repulsive interaction between anions *increases* the lattice energy. Moreover, if the anions are packed in such a way that the cation is rattling inside a void formed by the anions, this effect also increases the lattice energy. Collectively, to reach a balance among these competing effects and rationalize the ionic crystal structure, Linus Pauling proposed his famous five principles, also known as the 'Pauling rules' [3,4].

| Radius Ratio ($r_C/r_A$) | Coordination number | Type of geometry | |
|---|---|---|---|
| < 0.155 | 2 | Linear | 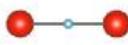 |
| 0.155 - 0.225 | 3 | Triangular | 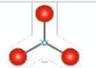 |
| 0.225 - 0.414 | 4 | Tetrahedral | 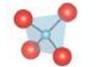 |
| 0.414 - 0.732 | 6 | Octahedral | 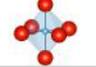 |
| 0.732 - 1.000 | 8 | Cubic | 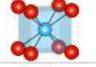 |
| 1.000 | 12 | Cuboctahedral | 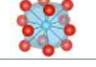 |

**Fig. 1:** *Coordination number and type of polyhedral geometry determined by the cation-anion radius ratio in TMOs. The central cation is displayed with a small blue cycle, the oxygen ions with a big red cycle. The black rod between each cation and oxygen represents ionic bonding.*

**Can we predict a crystal structure?** According to the Pauling rules, the structure of a complex ionic crystal is mainly controlled by two factors: the local coordination number (CN) together with the polyhedral geometry of a cation and the network of the polyhedra spanning the crystal. First, Pauling's rule determines CN and its polyhedral geometry: this is the cation-anion radius ratio rule. Figure 1 summarizes the typical CN values and the corresponding polyhedrons.

**Why is coordination so important?** After you learned about the local polyhedra of cations, the next step is to understand how these polyhedra are interconnected. In real TMO solids, three common polyhedral networks can exist: *corner*-sharing polyhedra, *edge*-sharing polyhedra, and *face*-sharing polyhedra. An important principle pointed out by Pauling is that *sharing of edges and especially faces by two polyhedra cost more energy than sharing corners*. This is because in edge-sharing and face-sharing cases, the cations are located in closer proximity, increasing the electrostatic repulsion among them. In addition, for TMOs with multiple cations, those of high chemical valency and small coordination numbers tend not to share polyhedron elements, increasing their distance and thus reducing the repulsive interaction between them.

At this point, let me introduce two popular TM compounds to make the discussion more concrete. *Perovskites $ABO_3$*. The perovskite structure is relatively simple and common for compounds with the chemical formula $ABO_3$. Here we find two alternative combinations of A and B cations. If the A site is a *rare-earth* ion and the B site is a *transition metal* ion (e.g., $RENiO_3$ with RE = La to Yb), the charge state of each ion is $A^{3+}B^{3+}O_3^{2-}$. Alternatively, if the A site is an *alkaline-earth* ion (e.g., $ATiO_3$ with A = Mg to Ba) and the B site is a transition metal ion, the charge state is given by $A^{2+}B^{4+}O_3^{2-}$. No matter what combination, the A ion must be larger than the transition metal B ion, and it should be coordinated by *twelve* oxygens. At the same time, the B transition-metal ion is surrounded by *six* oxygens forming the octahedral coordination, and the network of corner-sharing B octahedra is the hallmark motif for perovskites (see Fig. 2c). Thus, a perovskite's ideal conventional unit cell is cubic with a B–O–B bond angle of 90º. Of course, depending on the relative size of different ions, an actual unit cell can deviate from the cubic structure and it usually stabilizes in a lower-symmetry lattice. To predict if the structure deviates from the ideal cubic, in 1926, Goldschmidt introduced an index called *the tolerance factor* $t_G$, to quantify distortions in a perovskite crystal and predict the possible structure $t_G = (r_A + r_O)/\sqrt{2}(r_B + r_O)$, where $r$ is an ionic radius.

*Spinels* $AB_2O_4$. Compared to perovskite, the spinel structure is more complicated (see Figs. 2c and 8b). The general chemical formula for the spinel structure is $A_{1-\delta}B_\delta[A_\delta B_{2-\delta}]O_4$. When $\delta$ is 0, it is known as the *normal spinel* $AB_2O_4$, in which all A cations are tetrahedrally coordinated, while all B cations are octahedrally coordinated. When $\delta$ equals to 1, the chemical formula is $B[AB]O_4$, and known as the *inverse* spinel. In this case, the tetrahedral sites are occupied by half of the B cations, while the other half of the B cations and all the A cations occupy the octahedral sites. Finally, when $0 \leq \delta \leq 1$, A and B cations mix up in both the tetrahedral and the octahedral sites. Unlike perovskite which assume many lattice structures and space groups, spinels usually stabilize into a *cubic structure* (with space group Fd3m). As for the charge state, assuming $O^{2-}$ as is predominantly true in TMOs, there are two allowed ionic charge patterns for A and B ions: $A^{2+}/B^{3+}$ (common in almost all cases) or $A^{4+}/B^{4+}$ (rare but does exist, e.g., in $GeZn_2O_4$).

**How do electrons behave inside TMOs?** TMOs have a vast range of electronic behaviors, including those found in conventional metals and insulators, which are classified according to band theory. In addition, you can find various exotic phases such as high-temperature superconductivity, correlations-driven metal-insulator or Mott transitions, and topological states of quantum matter. The very diversity of TMOs makes them almost impossible to fit into a universal theory of their electronic properties. The usual approach in physics is to focus on the dominant term in the Hamiltonian while treating other terms as corrections or perturbations. Following this logic, in practice, the challenge of describing the electronic properties lies in selecting a starting point: whether the electrons are localized about corresponding ions or itinerant over the whole solid. Starting from those two extremes, several theoretical models have been developed, which I briefly introduced below (also see [5]).

*Ionic Model.* This simple but powerful model treats electrons from a local point of view. When we place a transition metal cation inside a solid, besides the Madelung potential, valence electrons of this cation experience additional Coulomb interactions stemming from surrounding oxygens, which we call the *crystal field*. Serving as a perturbation source, the degenerate energy levels obtained for isolated atoms are now split. Since the $d$ sub-shell is the outmost shell of a transition metal ion, the crystal field can significantly affect its energy. As a result, the five spherical harmonics labeled by their quantum numbers $(n, l, m)$ are no longer the eigenfunctions in the presence of a crystal field. Instead, we introduce new eigenfunctions that are linear combinations of those spherical harmonics.

These eigenfunctions' shapes (or electron density distributions) are plotted in Fig. 2a. Figure 2b displays a few distinct $d$ orbital energy splitting patterns under different crystal field symmetry. Depending on the crystal field's local symmetry or the polyhedral coordination's geometry, the splitting sequences can be quite different. In this lecture, I will mainly discuss two types of crystal fields: octahedral and tetrahedral. In the octahedral field, the orbitals are split into an upper doublet $e_g$ group including $d_{x^2-y^2}$ and $d_{3z^2-r^2}$ orbital states and a lower triplet $t_{2g}$ group with $d_{xy}$, $d_{yz}$, and $d_{xz}$ states.

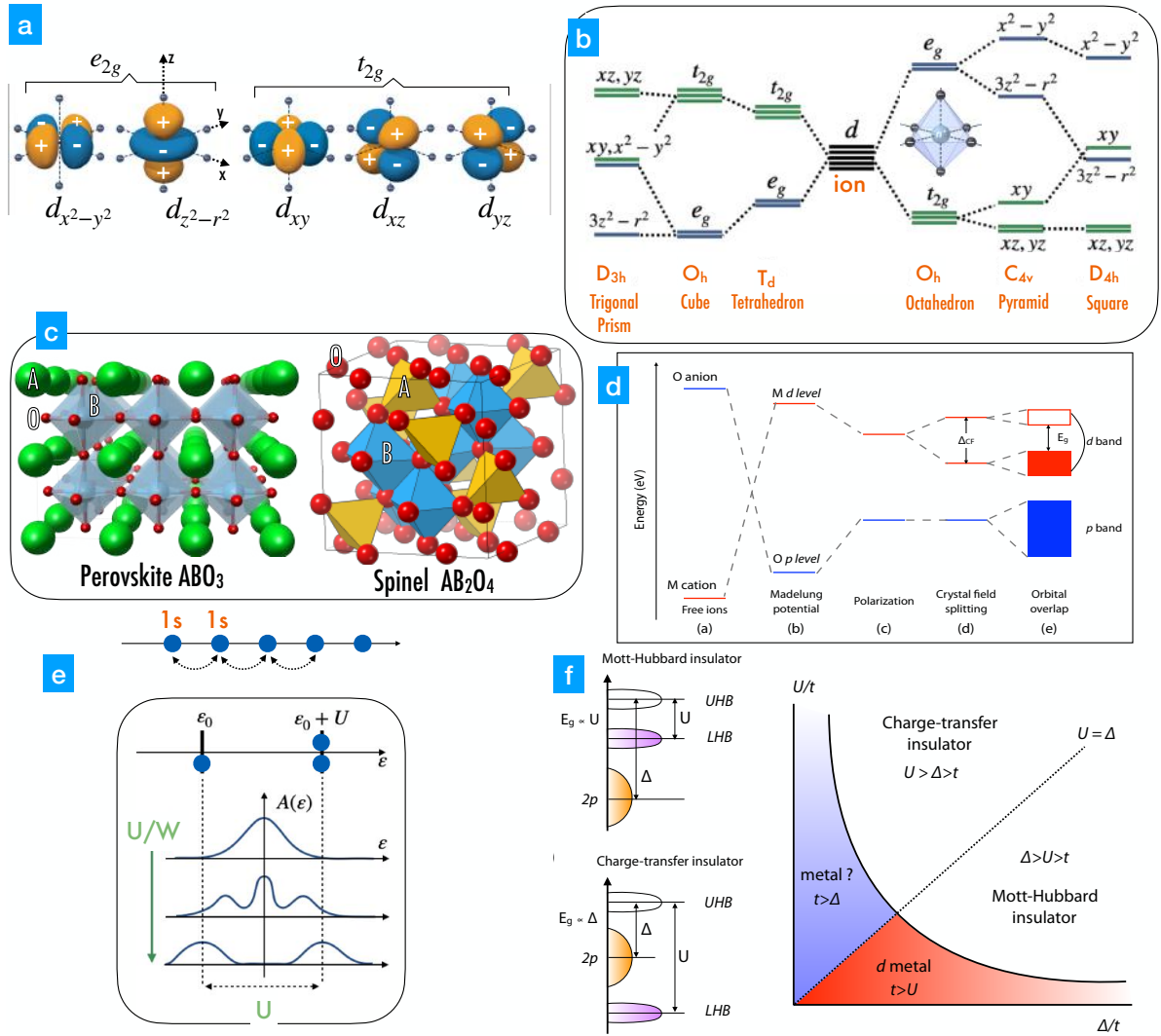

**Fig. 2:** *a: Shapes of the $d$ orbitals. The notion of $x$, $y$, and $z$ refer to the wave-function variables written in the Cartesian coordinate system. b: Energy splitting of the $d$ orbitals under different crystal field symmetry. The notation for $e_g$, $t_{2g}$ are defined according to group theory. c: Perovskite $ABO_3$ lattice and $AB_2O_4$ spinel unit cell. d: (a)–(e) Evolution of the electronic energy levels of TMOs in the ionic model. O and M refer to oxygen and transition metal ion, respectively. e: A schematic of the $d$ band in the Hubbard model. For a half-filled band, the electron correlations (Hubbard $U$) are able to open a gap when its strength reaches the critical value, resulting in the lower (UHB) and upper Hubbard (LHB) bands separated by the correlated gap of $U$. f: (left top) Energy level diagram of a standard Mott-Hubbard insulator (MHI). In this case, the gap $E_g$ is defined by $U$. (left bottom) Energy level diagram of the charge-transfer insulator (CTI). In this case, the gap $E_g$ is defined by the charge-transfer $\Delta$ energy and intimately involves states on oxygen or anion in general. (right) Zaanen-Sawatzky-Allen phase diagram. Note, $U$ and $\Delta$ represent the electron-electron correlation and charge-transfer energy, respectively. $t$ is the electron hopping strength, a measure of the electronic bandwidth.*

How can we intuitively understand the origin of the energy splitting without relying on a group theory analysis? Recall that oxygen $p$ orbitals are dumbbell-shaped and point along the Cartesian axes. In an octahedral environment, the lobes of $e_g$ orbitals always point towards those of $p$ orbitals, effectively increasing the Coulomb repulsions between $e_g$ and $p$ electrons and thus lifting up the energy levels. In contrast, $t_{2g}$ orbitals have lobes pointing away from those of oxygen. Therefore, all $t_{2g}$ levels are shifted downward. The energy gap between these two groups is denoted as $\Delta_{CF}$ (or 10Dq in chemistry). A similar analysis can be applied in the case of a tetrahedral environment, where the *splitting pattern is reversed* compared to that of octahedral coordination.

*Hubbard model.* The combination of the ionic model with the traditional band theory can successfully describe the electronic structure of many TMOs reasonably well. However, as early as 1937, it has been recognized that several transition metal compounds (e.g., CoO, NiO, $Fe_2O_3$), which are expected to be metallic, instead are wide-gap insulators. This failure in predicting the ground state in these materials signals that some critical factors are missing. The main reason is that in the ionic model, the electrons are considered independent and Coulomb repulsion is therefore omitted. This interaction, also known as electron correlation, is weak when electrons move in broad bands. However, in TMOs, the partially filled bands derived from $d$-electrons are usually very narrow, and the electrons appear more localized. Under this circumstance, the electron-electron correlations are inevitably amplified, exerting significant influence on the band structures and the overall physical properties of the materials. To quantitatively account for this observation, in 1963 Martin Gutzwiller, Junjiro Kanamori and John Hubbard independently proposed a new model Hamiltonian $\hat{H} = -t \sum_{\langle I,j \rangle, \sigma}(c_{i,\sigma}^\dagger c_{j,\sigma} + c_{j,\sigma}^\dagger c_{i,\sigma}) + U \sum_i n_{i,\uparrow} n_{i,\downarrow}$ [6–8].
In this Hubbard Hamiltonian, the first term describes the usual hopping effect of electrons from a site to its nearest-neighbors without spin-flip (so called kinetic term), whereas the second term accounts for the extra repulsive energy cost due to double occupation of the same lattice site. In this sense, the Hubbard model includes two competing processes (localization vs. delocalization), and the true ground state is determined by the relative strength between Hubbard $U$ and hopping integral $t$, which is proportional to the electronic bandwidth $W$. The influence of Hubbard $U$ on the electronic band structure is shown in Fig. 2e.

*Mott insulators and Mott transitions.* Now let us recap that in accordance with the original band theory, if a valence band with $2N$ capacity is half filled, the system is a metal. Nevertheless, now electron-electron correlations act to open a correlated or Coulomb gap in the valence band. If the correlation effect is weak compared to the bandwidth ($U/t \ll 1$), the band will not be split, and the material remains metallic. However, if the correlations are strong or $U/t \gg 1$, a correlated band-gap emerges and separates the $2N$ valence band into two new bands, called upper Hubbard band (UHB) and lower Hubbard band (LHB), each with $N$ electron capacity [9]. For this reason, the material turned into an insulator, is collectively known as a Mott insulator or Mott-Hubbard insulator. I must stress that the Mott insulator is a highly non-trivial state, and many of the transition metal insulating compounds, which are predicted to be metallic, belong to this new class of quantum materials. It is also helpful to remember that for trivial band insulators or semiconductors, the energy gap is defined by the periodic potential of the crystal lattice. In sharp contrast, in Mott crystals, the energy gap arises solely from electron-electron correlations.

Let me dig deeper into the excitation spectrum of Mott compounds. Naturally, starting from the noninteracting side ($U=0$) and gradually increasing $U$, we should sooner or later reach a critical point $U_{cr}$ where a metal-to-insulator transition takes place. This transition is called the Mott transition. Among TMOs, a significant number of compounds undergo such a transition. It has been observed that several factors, such as temperature, pressure, and electric and magnetic field, can trigger the Mott transition. Another important fact is that even though the Hubbard Hamiltonian is not explicit about the underlying crystal structure, this Mott transition is usually accompanied by structural distortions and long-range spin orderings. Because of the entwined couplings, the question of a driving force behind the Mott transition, including the role of electronic correlations, is still not entirely understood.

*Charge transfer insulators.* Up to this point, I have considered only the effects of transition metal $d$ bands. You should remember that anions and, specifically, oxygens are also very important. In fact, in many TMOs, the oxygen $2p$ bands are slightly lower in energy than the $d$ bands. Here we can also ask, once the UHB and LHB are formed, what are their relative positions with respect to the oxygen $p$ bands? As illustrated in Fig. 2f, there are mainly two cases of energy level diagrams expressing their relative positions. The oxygen $p$-band can either be lower than both of the Hubbard bands or in between these two bands. TMOs in the former case are a standard Mott insulator, whereas those in the latter case are given a new name, a change-transfer insulator (for an excellent discussion see [10, 11]).

To explain the difference, it is necessary to introduce a new energy scale, the so-called charge-transfer energy $\Delta_{CT}$. As there are $n$ electrons in the $d$ levels, $d^n$ configuration, two types of excitations exist. First, the electron can either hop onto another already occupied site in the same $d$ level, say $d^n d^n \leftrightarrow d^{n-1} d^{n+1}$, or an electron from oxygen $2p$ band can hop onto the empty $d$-state, $d^n p^6 \leftrightarrow d^{n+1} p^5$. The first process costs us the energy $U$, while the second one cost a certain amount of charge-transfer energy, which is $\Delta_{CT} = E_d - E_p$.

The lowest charge excited state can be different depending on the ratio of $U$ to $\Delta_{CT}$. When $\Delta_{CT} \geq U$ [see Fig. 2f(left)], the oxygen $p$ band lies lower than the LHB. The band gap is now determined by $U$. $d^n d^n \leftrightarrow d^{n-1} d^{n+1}$ costs less energy, as is typically the case for a conventional Mott insulator. However, when $\Delta_{CT} \leq U$, the electron hopping between oxygen and TM ion $d^n p^6 \leftrightarrow d^{n+1} p^5$ costs less energy and results in the lowest excited state. Here, the band gap is defined by the charge-transfer energy $\Delta_{CT}$ and not by $U$. The related materials are called charge-transfer insulators; in this case, the excited electrons are from the oxygen $p$ levels. Moreover, it turns out that many of the physically interesting TMOs belong not to the Mott but to the charge-transfer family; the prototypical examples are the high-temperature superconducting cuprates, e.g., $La_2CuO_4$ or $YBa_2Cu_3O_{7-\delta}$.

Even more complicated situations occur if you consider the bandwidth, which is proportional to the electron hopping strength $t$. Specifically, what I considered so far was based on the assumption that $\Delta_{CT} \gg t$ and also $U \gg t$ so that one can expect the insulating ground state to emerge. However, you can imagine that if the bandwidth is large enough, the correlated gap may fail to open, resulting in a metallic ground state. These ideas are rationalized in the so-called Zaanen-Sawatzky-Allen (ZSA) phase diagram, which is shown in Fig. 2f(right). For most transition metal compounds, their electronic structure can be qualitatively explained by this phase diagram. One significant effect is still missing from my discussion – the spin-orbit effect. The subject is so vast and important that instead, I refer the reader to the reviews [9, 12].

## 2  What are correlated oxide interfaces?

One of the prime goals of basic condensed matter physics is to seek out and explore new collective quantum states. Towards this goal, ultra-thin heterostructures composed of two or more structurally, chemically, and electronically dissimilar constituent oxides have been developed into a powerful approach over the past few decades [13–18].

Here, the main idea is that at the interface where the dissimilarities meet, the frustration caused by mismatches between the arrangement of atoms, charges, orbitals, or spins can trigger the emergence of phenomena with electronic and magnetic properties which non-trivially differ from the bulk compositions. For many research groups, the correlated interface engineering has opened a route to new materials behaviors using those mismatches as the control parameters. Paraphrasing the Nobel Prize winner Herbert Kroemer, 'The interface is a new material.'

A summary of the potential mismatches at oxide interfaces is shown in Fig. 3(left). As seen, at the oxide interfaces, the following degrees of freedom, can be rationally designed:

(1) *Epitaxial strain*. Strain results from lattice mismatches between the atomic arrangement of two different TMOs. By delicately applying strain, the M-O-M bond length and bond angle can be effectively tuned, which in turn may trigger electronic and magnetic phase transitions. For instance, using epitaxial strain as the control parameter, ultra-thin $NdNiO_3$ films have supported a remarkably enlarged phase diagram with several new states not observed in bulk [19];

(2) *Local symmetry*. As discussed in the previous section, TMOs have a variety of local symmetry or coordinate polyhedrons. Suppose we want two components with different local symmetries to grow as a heterostructure. To achieve growth, each subsystem needs to compromise, and the interfacial structure will deviate dramatically from their bulk counterparts. A robust example is the $\gamma$-$Al_2O_3$/$SrTiO_3$ interface: $\gamma$-$Al_2O_3$ has a spinel structure with tetrahedral and octahedral local symmetries, whereas $SrTiO_3$ is a perovskite with only octahedral symmetry. However, at the interface, an anomalous *square pyramid local symmetry* emerges for Ti ions, which strongly alters the electronic band structures;

(3) *Polar mismatch*. Even with the same lattice structure, the net charge of each atomic layer can be distinct from the naive ionic pictures. If a heterointerface consists of entities with different net charges per atomic plane, in this case the so-called 'polar mismatch' may occur, and the charges near the interface must rearrange to satisfy a condition of charge neutrality. This phenomenon has been found in many complex oxide heterostructures and for example was conjectured to be the source of the two-dimensional electron gases (2DEG) emerging between two *insulating* TMOs [14];

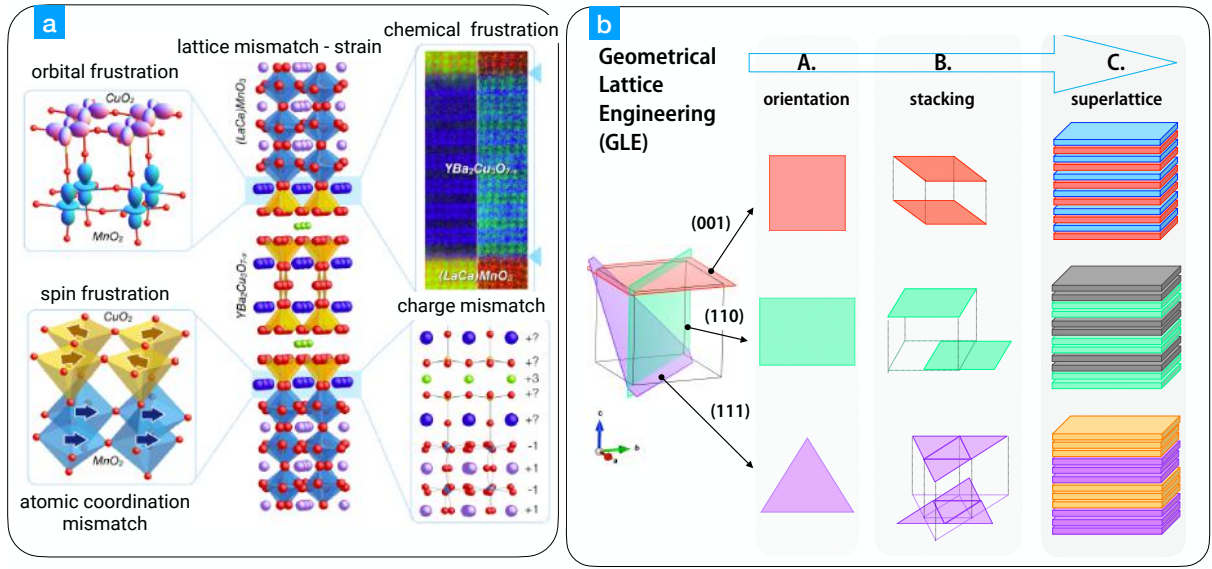

**Fig. 3:** *a: Mismatches at the correlated interface. b: Schematic illustration of the idea about geometrical lattice engineering, which has three control parameters named as orientation, stacking, and superlattice throughout its entire process.*

(4) *Orbital reconstruction*. Quite often, we find a variation of the orbital occupation at oxide interfaces that stems from the modulations of the atomic structure or charges due to the mismatches. As a result, unusual orbital configurations can be realized at the interface. To illustrate, in ultra-thin $LaNiO_3/LaAlO_3$ superlattices, it was found that unlike the bulk $LaNiO_3$ ($Ni^{3+}$, $3d^7$), where valence electron equally occupies the $d_{x2-y2}$ and $d_{z2}$ orbitals, at the superlattices interfaces $d_{x2-y2}$ is preferentially occupied [20];

(5) *Magnetic coupling*. Oxide interfaces can be an effective tool to tune or even design magnetic interactions. You can imagine that if materials with a different kind of exchange coupling (for instance, FM vs. AFM) are attached in atomic proximity, their incompatible order parameters may eventually drive them to form a new magnetically balanced state (e.g., helical or canted spin arrangements). For example, an interesting phenomenon has been discovered at the interface of a high-$T_c$ superconducting cuprate with a colossal magnetoresistance manganite, $YBa_2Cu_3O_7/La_{2/3}Ca_{1/3}MnO_3$, where surprisingly superconductivity and ferromagnetism coexist [21].

## 3 New quantum materials by geometrical lattice engineering

Inspired by the success of those interface engineering methods, recently another promising method, collectively known as "geometrical lattice engineering" (GLE), has been presented as a powerful tool to forge new topological and quantum many-body states. In close synergy with interface and strain engineering, where mismatches between layers induce unusual interactions, the key idea behind the GLE is to design fully epitaxial ultra-thin films and heterostructures with an *artificial lattice geometry generated by stacking of a precise number of atomic planes along a specific orientation* [22].

This concept can be illustrated by recognizing that the properties of a three-dimensional (3D) material can be drastically altered by changing parameters such as: the stacking of two-dimensional (2D) atomic planes, the specific arrangement of ions in those planes, their sequence, and the periodicity of layers fulfilling the charge neutrality condition. Conventionally, for thick bulk-like films, the effect of those variations is often negligible (maybe apart from anisotropy). In sharp contrast, for ultra-thin films, the directional stacking of atomic planes becomes dominant in defining the electronic and magnetic properties. Following this idea, many exciting materials systems have been theoretically proposed in pursuing exotic quantum states. At the same time, the experimental work on GLE has been primarily focused on the growth of cubic or pseudocubic (111)-oriented artificial lattices. In general, throughout the process of heteroepitaxial fabrication, to be able to design a new material by the GLE, you can follow three controllable routes. To explain this further, I will use a 3D simple cubic unit cell model to illustrate those control parameters (see Fig. 3b).

*Growth orientation.* Starting with the same bulk compound, its 3D crystal structure can be viewed as atomic layers stacking with different in-plane lattice geometries along different crystallographic directions. For example, as illustrated in Fig. 3 (right), while the (001) planes have square symmetry, the (110) and (111) planes have rectangular and triangular geometries, respectively. The required in-plane lattice geometry by design can be determined by selecting a proper structure and orientation of the substrate surface, which acts as a guiding template during the initial nucleation and growth stages. A typical example is the realization of a 2D magnetic lattice with extreme frustration derived from the ultra-thin (111)-oriented spinel-type structure $AB_2O_4$. This example I will describe in detail in Sec. 5.5.

*Out-of-plane stacking sequence.* In bulk crystals, the periodicity of the atomic planes can vary dramatically based on the choice of crystallographic direction to fulfill the requirement of translational symmetry and the relative atomic positions of neighboring lattice planes . For instance, the stacking of the adjacent layers can be either right on top of each other [the (001) stacking in Fig. 3], or shifted [the (110) stacking], or even entirely reversed [the (111) stacking]. This observation is at the heart of the design of artificial heterostructures since by controlling the number of stacking planes within that period you can forge unique quasi-2D lattices that do not exist in the naturally formed crystals. Among the prominent examples of GLE, I want to mention the generalized graphene lattice, which can be obtained by digitally tuning the number of atomic layers of (111)-oriented $ABO_3$ perovskite-type structures. I will present this case later in Sections 5.3 and 5.4.

*Isostructural superlattices approach.* Combining isostructural materials to establish superlattice structures with digital control over the individual number of layers adds another practical dimension to applying GLE. This approach can be very useful for achieving materials with complex chemical compositions or even thermodynamically unstable phases in the bulk form. A representative test case is the fabrication of (111)-oriented $1ABO_3/1AB'O_3$ superlattices [here "1" refers to a single cubic (or pseudocubic) unit cell] that gives rise to an $A_2 BB'O_6$ double perovskite [23].

# 4 How can we grow perfect interfaces?

In Section 1 we briefly discussed many theoretical concepts; now it is time to turn to something more applied. In this section, I want to focus on the question: 'How can we grow multi-layer structures with high-quality interfaces to match existing theoretical proposals?' Here, I describe one of the most popular methods for synthesizing such artificial complex oxide structures, called pulsed laser deposition or PLD. Despite its relatively young age, PLD has proven a versatile method for fabricating exceptionally high-quality epitaxial thin films and heterostructures during the last two decades (see [24] and the comprehensive references [25, 26]).

Compared with other popular physical vapor deposition (PDV) techniques, such as magnetron sputtering or all-oxide molecular beam epitaxy, several advantages make PLD particularly successful in growing complex oxide films. These include modestly priced instrumentation, stoichiometric transfer of ions from targets onto a substrate, an energetic forward-directed plume, and hyper-thermal interaction of the ablated species with the background gas (e.g., oxygen, nitrogen, argon). In other words, it is a PVD process performed in a high vacuum or low-pressure system using a pulsed laser as the heating source of ionic and molecular species.

What does a typical growth cycle look like? During the deposition, a pulsed laser with a pulse duration of $\sim$ 20 ns operating in the UV spectrum ($\lambda = 248$ nm) is focused on a small portion of a ceramic/polycrystalline target, which usually contains a stoichiometrically correct mixture of atoms to be synthesized on the substrate as the desired film. With a sufficiently high energy fluence of 1–2 J/cm$^2$, the ejected ions/molecules from the target vaporized by each laser pulse produce a directional plasma plume. Next, this highly forward-directed plume moves towards the substrate in a background gas atmosphere ranging from the ultra-high vacuum of $10^{-8}$ Torr and up to 1 Torr. This flux of oxidized and cooled to thermal energies ionic/molecular species rapidly propagate, reach the substrate, and eventually nucleate and crystallize into atomic layers of epitaxial solid films. To make a high-quality structure, you need to optimize several control parameters. Let me start with the targets. Since a complex oxide compound typically contains two or more kinds of atoms, a solid target suitable for PLD should be uniform and highly dense, possessing identical cation stoichiometry to the desired film. If the laser ablates a loose target, the resultant film will have a rough surface with microscopic molten droplets ejected from the target. The substrate crystal should closely match the lattice parameter and symmetry of the desired film. Finally, the crystallographic orientation of the substrate surface is critical as it determines the epitaxial orientation of the film; the substrate serves as the atomic template during the nucleation and the initial stage of film growth.

Once you select a substrate and the target, the remaining factors affecting the deposition process are laser energy and fluency, the distance between the target to the substrate, background gas pressure, and substrate temperature. As for the laser fluency, if we set it too low, the result of each laser pulse would be similar to thermal heating. In this case, the ejected flux of ionized species may deviate from the desired stoichiometry of the target due to the differences in vapor pressure among each constituent chemical element. To avoid this issue, we need to increase the laser fluence high enough to overcome the ablation threshold for a specific target, above which the evaporation is independent of the vapor pressure, and the plume can maintain its proper stoichiometry. At the other extreme, running deposition with too high fluence would render the formation of macroscopic droplets or even damage the target. As a result, the typical laser fluence range is set around 1–4 J/cm$^2$. Next, to grow complex oxide films, molecular oxygen is often introduced into the chamber as the background gas for two reasons. First, the ejected low-mass molecular species needs to interact with $O_2$ to get the desired phase (remember, there is no solid oxygen!). Secondly, the background gas is required to reduce the high kinetic energies of the plume from several tens of eV down to meV; without a reduction of the kinetic energy, the complex ions of the plume would collide with the substrate's surface, potentially sputtering off the newly created island of crystalline phase and/or creating defects.

The substrate temperature is another critical factor in determining the quality of the films. However, temperature's role during the deposition process is rather complicated. On the one hand, a high substrate temperature is usually favored since it enhances the mobility of adatoms so that they can rearrange, forming a flat surface morphology. On the other hand, high temperatures can evaporate constituents with high vapor pressures out of the film, resulting in oxygen vacancy defects or missing cations. Another issue occurs when growing a superlattice structure composed of various oxide components. The thermally active atoms of each constituent layer can diffuse across the interface, destroying the atomic sharpness of the interface. In addition, for many complex oxides with low crystal symmetry, the epitaxial orientation of the film is very sensitive to the substrate temperature, often leaving a relatively narrow window for each phase.

Finally, under typical growth conditions, the deposition rate varies from a few tenths to one angstrom per laser pulse. This feature ensures precise control of growth on the sub-monolayer level and makes PLD a good choice for fabricating multilayers and superlattices of complex multi-element materials.

I only have discussed the most general trends for each essential control parameter. The specific values of those parameters are truly material- and growth chamber-dependent. This means that for every new material of interest and for each specific growth machine, finding and optimizing a comprehensive phase diagram for the best growth condition is necessary.

# 5  Examples of correlated oxide interfaces

In what follows there are some interesting examples of recent correlated oxide interfaces. The prime criterion for such a selection is to present the reader with the synthetic quantum materials that harbor emergent states or phases not seen in the bulk constituent layers. Also, unlike the previous sections, this section is more technical because it necessarily relies on the application of multiple state-of-the-art probes and advanced computational methods; as such, I suggest to treat those examples as a show-and-tell.

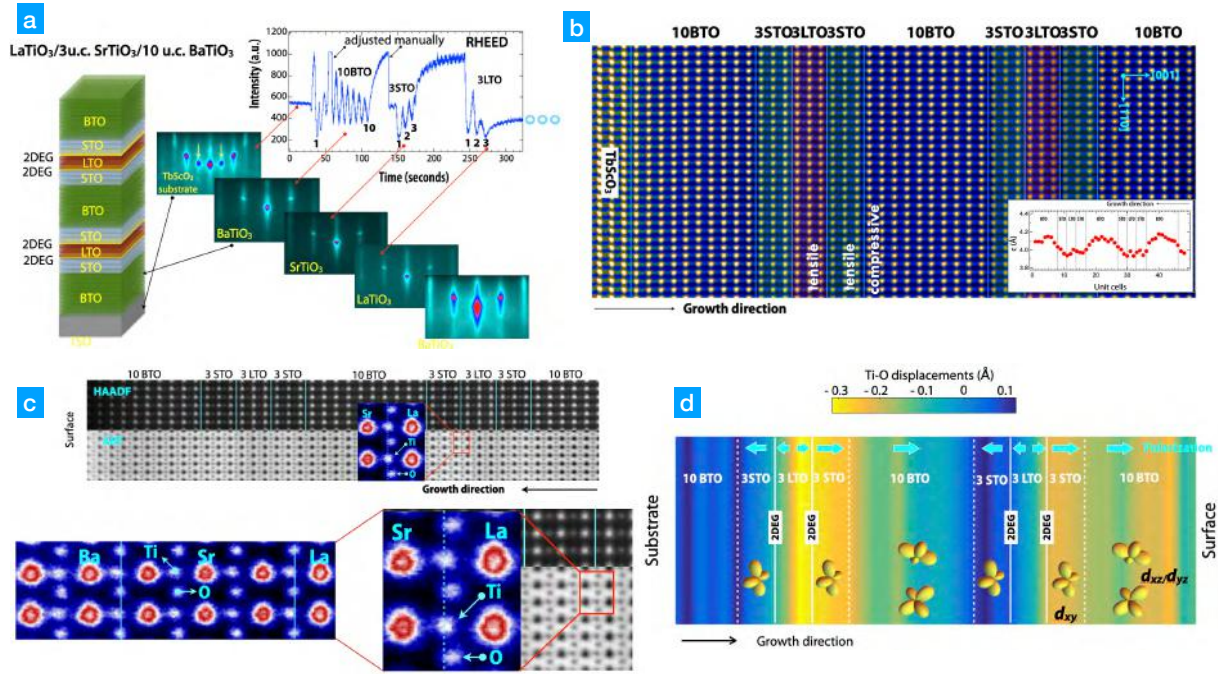

**Fig. 4:** *a: Diagram of a BTO/STO/LTO superlattice, where the yellow layer indicates the formation of a 2DEG. In-situ RHEED images confirm high crystallinity. b: STEM/EELS image reveals high-quality interfaces. Inset: extracted out-of-plane lattice parameters for each individual layer. c: High-resolution STEM imaging reveals significant Ti-O polar displacements, resulting in 8% enlargement of the lattice parameters. d: Summary of strength and direction of polarization.*

## 5.1 Artificial ferroelectric metal

Polar metals, commonly defined by the coexistence of polar crystal structure and metallicity, are thought to be scarce [27, 28], because long-range electrostatic fields favoring the polar structure are expected to be fully screened by the conduction electrons of a metal. Generally, based on the type of atomic displacements, polar metals with perovskite structure fall into two main categories: A-site driven (e.g., positional shifts of Li, Nd, and Ca ions in $LiOsO_3$ [29], $NdNiO_3$ [30], and $CaTiO_{3-\delta}$ [28], respectively) or B-site driven (e.g., a shift of Ti ions in $BaTiO_{3-\delta}$ [31, 32]) kinds. For the former category, recent theoretical work [28] has suggested the absence of a *fundamental incompatibility between the polarity and metallicity*, whereas, for the latter, polar displacements show a rapid decrease with increasing carrier concentration [32].

Moreover, in reducing from three to two dimensions, it remains an open question whether a polar metal can exist at all. Here, I describe the realization of a room temperature two-dimensional (2D) polar metal of the B-site type in *tri-color* (tri-layer) superlattices $BaTiO_3/SrTiO_3/LaTiO_3$. Such an explicitly non-centrosymmetric 2D metal provides a template to engineer an interesting quantum many-body state with three coexisting phases – ferroelectricity, ferromagnetism, and superconductivity.

Let us start with the design idea. First, we use tri-color rare-earth titanate heterostructures (see Fig. 4a) made of a layered arrangement of the ferroelectric alkaline-earth titanate $BaTiO_3$ (BTO, green), the paraelectric alkaline-earth titanate $SrTiO_3$ (STO, gray), and the Mott insulator rare-earth titanate LaTiO$_3$ (LTO, red). Bulk BaTi$^{4+}$O$_3$ with $3d^0$ electron configuration is a well-known ferroelectric material, which can undergo complex structural and ferroelectric phase transitions on cooling, e.g., from cubic to tetragonal near 400 K, tetragonal to orthorhombic near 280 K, and orthorhombic to rhombohedral near 210 K [33] (ferroelectric properties are present in the latter three phases). Bulk SrTi$^{4+}$O$_3$ with $3d^0$ electron configuration is a band insulator with a charge gap size of $\sim 3.3$ eV. In contrast, bulk LaTi$^{3+}$O$_3$ with $3d^1$ electron configuration is a Mott insulator and undergoes a G-type antiferromagnetic transition below 146 K [34]. In bulk, the lattice parameters are a = 3.905 Å for cubic SrTiO$_3$; 4.00 Å for cubic BaTiO$_3$; 3.958 Å for pseudocubic TbScO$_3$; and 3.956 Å for pseudocubic LaTiO$_3$. Based on these lattice parameters, the SrTiO$_3$ layers of BTO/STO/LTO superlattices grown on a TSO substrate are under tensile strain, whereas the BaTiO$_3$ layers are under compressive strain.

Another interesting feature of these designer superlattices is in the transfer of electrons from LTO ($3d^1$) into the STO ($3d^0$) layers leading to the formation of a two-dimensional electron gas (2DEG) [35] at the interfaces (yellow layer in Fig. 4a), which have a shared polar structure due to the presence of ferroelectric BTO. As clearly seen, this design has two inequivalent interfaces, BTO/STO and STO/LTO. What is interesting is the fact that both bi-color BTO/STO and BTO/LTO interfaces are insulating. Therefore, the metallicity in the 3-color BTO/STO/LTO structure comes from the 2DEG formed at the vicinity of the STO/LTO interface alone. Also, in contrast to itinerant electrons of the STO/LTO interface, the electrons at BTO/LTO interfaces are still localized, forming no 2DEG.

To monitor the crystal structures of the BTO/STO/LTO superlattice during growth, the measurements of in-situ reflection high-energy electron diffraction (RHEED) were performed. As seen in Fig.4a ultra-thin tri-color superlattices consisting of (BTO)$_{10}$/(STO)$_3$/(LTO)$_3$ (here the subscript refers to the number of unit cells) as well as reference samples of (BTO)$_{10}$/(LTO)$_3$ superlattice and BTO thin film were synthesized on TbScO$_3$ (110) single-crystal substrates by pulsed layer deposition in a layer-by-layer mode. During growth, the RHEED diffraction patterns for substrates, BTO, STO, and LTO layers stayed sharp, confirming the high crystallinity and epitaxy of BTO/STO/LTO superlattices.

Further, to determine the atomic-scale structures of the superlattices, their interfacial structure, and their chemical composition, the authors applied cross-sectional scanning transmission electron microscopy (STEM) with electron energy-loss spectroscopy (EELS). Figure 4b shows the tri-color superlattice's high-angle annular dark-field (HAADF) STEM image, revealing high-quality continuous and coherent interfaces without phase separation. In the Z-contrast HAADF image, the expected layer thickness and designed sequence of three layers are clearly distinguishable from the different intensities due to the difference in the scattering power of the layers. Additionally, as seen in the inset (red dots) in Fig. 4b, the periodicity of the growth sequence was determined from the periodic variation of the extracted out-of-plane lattice parameters for individual BTO, STO, and LTO layers. As engineered for the interfacial charge transfer, low-temperature electrical transport measurements of BTO/STO/LTO revealed the expected metallicity and large carrier density of conduction electrons, $n_c$ in all tri-color samples (n$_c \sim 10^{14}$ cm$^{-2}$).

Next, I want to discuss the microscopic details of the centro-symmetric breaking of TiO$_6$ octahedra leading to the formation of a 2D polar metal in this structure. To address this, high-resolution HAADF- and ABF-STEM imaging were carried out, which allowed for the direct observation and extraction of precise atomic positions of all constituent atoms, including oxygen, across the interfaces. As shown in Fig. 4c, significant Ti-O polar displacements, i.e., relative shifts of Ti and O along the out-of-plane direction, are found in the BTO/STO/LTO tri-color structure. Additionally, a detailed quantitative analysis of the ABF-STEM image was performed to determine the amplitudes and directions of the *polar displacements*. The Ti-O polar displacements are found to be as large as 0.3 Å (!), which is an almost 8% enlargement of the lattice parameters. Moreover, these large Ti-O polar displacements not only exist in BTO but also extend deep into the STO and LTO layers, where the 2D metallic layer resides, thus inducing polar displacements into the metallic interface.

*What about microscopic polarization and the connection to orbitals?* Figure 4d summarizes the strength and direction of polarization labeled by the color map from blue to yellow. A striking feature is the periodic reversal of polar directions across the Mott LTO layers. It can be attributed to atomic displacements driven by local up-down symmetry breaking, typical of perovskite surfaces, at the STO/LTO interface [36]. More specifically, as seen in Fig. 4d, the authors find that around the LTO region, $d_{xy}$ states are predominantly occupied. This orbital polarization decays exponentially. However, in the BTO region, $d_{xz}/d_{yz}$ states are mainly occupied with the density shifted towards the left BTO/STO interface. The spatial separation of $d_{xy}$ and $d_{xz}/d_{yz}$ states is the combined effect of the electrostatic energy and the crystal field splitting [37]; namely, in the LTO region, the electrostatic potential from positively charged (LaO)$^{1+}$ layers dominates and is screened by $d_{xy}$ electrons having in-plane dispersion. However, in the BTO region, the out-of-plane (or apical) Ti-O distances become substantially larger compared to the in-plane Ti-O distances due to the elongated c-lattice constant. This, in turn, lowers the on-site energy of $d_{xz}/d_{yz}$ orbitals and results in the large increase in the $d_{xz}/d_{yz}$ orbital occupancy compared to STO/LTO heterostructures.

## 5.2 Orbital assisted Kondo lattice and spin-polarized 2D electron gas

Magnetic interactions between the localized spins and conduction electrons are fundamental in many quantum many-body effects. Phenomenologically, in materials with localized spins coupled to conduction electrons, the Kondo interaction [38–40] competes with the magnetic Rudderman-Kittel-Katsuya-Yosida (RKKY) interaction [41], conceptualized in the so-called Doniach phase diagram [42] and Kondo lattice models [41]. In real transition metal crystals, however, the ground state depends on the strength of exchange interaction $J$, the electronic density ratio $n_m/n_c$ of the localized magnetic moments $n_m$ to conduction electrons $n_c$, and the orbital character of magnetically active electrons [43]. In the strong-coupling regime with large $|J|$, the Kondo interaction prevails and results in a Kondo singlet state [1], whereas on the weak-coupling side (small $|J|$), depending on the ratio $n_m/n_c$ [44], the RKKY interaction may give rise to either a ferromagnetic (FM) or antiferromagnetic (AFM) order between the localized spins. Notably, in the limiting case of $n_m/n_c \gg 1$, the localized spins tend to form ferromagnetic order by polarizing the conduction electrons [45]. In short, if we devise such a Kondo active structure with the specific set of control parameters described above, we should realize a highly desired artificial quantum material for spintronics with spin-polarized 2D metallicity.

In correlated $d$-electron heterointerfaces, the density ratio of $n_m/n_c$, the dimensionality, and the orbital polarization of the magnetic interactions are all vital components for the formation of a ground state [2]. Considering the splitting of the Ti $t_{2g}$ band between $d_{xy}$ and $d_{xz}/d_{yz}$ subbands is a prime cause for the interesting emergent phenomena in the SrTiO$_3$-based heterostructures [46]. This raises an important question: What is an experimental phase diagram for the emerging magnetic interactions at the nanoscale in the tri-color system?

To answer this question, a set of tri-layer (tri-color) superlattices composed of [3 u.c. LaTiO$_3$/$n$SrTiO$_3$/3 u.c. YTiO$_3$] ($n = 2, 3, 6$ unit cells or u.c.) and reference bi-layer samples [$m$LaTiO$_3$/$n$SrTiO$_3$] (here $m = 3, 20$ u.c. and $n = 2, 3, 6$ u.c.) and [3 u.c. YTiO3/$n$SrTiO$_3$] ($n = 2$ and 6 u.c.) were epitaxially synthesized by PLD on TbScO$_3$ (110) substrates. Again, in this designer system, the interfacial charge transfer is used to create a two-dimensional electron gas at the interface between LaTiO$_3$ and SrTiO$_3$ (LTO/STO) connected to a spatially separated interface with localized magnetic moments at the YTiO$_3$/SrTiO$_3$ interface (YTO/STO) (see Fig. 5b).

Before I discuss the ground state of such a tri-color structure, let us investigate each component of the superlattice. Figure 5a shows the magnetic phase diagram of $R$TiO$_3$ ($R$=La, Pr, Nd, Sm), which exhibits an antiferromagnetic (AFM)-to-ferromagnetic (FM) phase transition across the series of rare-earth titanates [47]. For our tri-color design, we select LaTiO$_3$ (∼0.2 eV gap, Ti $3d^1$) with the smallest distortion, which undergoes a G-type AFM phase transition below 146 K [34]. SrTiO$_3$ is a $3d^0$ system with no magnetism. And finally, for the third layer, the authors selected a significantly distorted YTiO$_3$ (∼1.2 eV gap, Ti $3d^1$), where the FM order forms below 30 K [48].

To summarize the design idea: In this structure, the action is supposed to happen at the LTO/STO and STO/YTO interfaces. The STO layer acts as an active spacer which depending on its thickness, brings the 2DEG from LTO/STO closer or further away from the magnetic moments of the STO/YTO interface. That is why I name this unique tri-color design the structure to explore *interacting in space order parameters*.

Now, what about the tri-color system? Figure 5b shows a high-angle annular dark-field (HAADF) STEM image of the tricolor superlattice, revealing high-quality coherent interfaces without phase separation. The atomic positions of the elements La (large yellow dots), Sr (large white dots), Y (large blue dots), and Ti (small green dots) are labeled schematically. To check for the presence of the interfacial charge transfer across the two LTO/STO and STO/YTO interfaces, the layer-resolved electronic structure of 3LTO/6STO/3YTO was investigated by atomic scale STEM-EELS line scanning across the interfaces. As shown in Fig. 5b, by scanning atomic layer-resolved Ti $L_{2,3}$-edge spectra across LTO/STO and STO/YTO interfaces [along the white-dashed line in Fig. 5b] with a high energy resolution of 0.4 eV and a high spatial resolution of 0.8 Å, the evolution of the Ti electronic structure across the interfaces was mapped out. The bottom panel of Fig. 5b shows the reference spectra for Ti$^{4+}$ and Ti$^{3+}$ acquired from bulk-like $SrTiO_3$ and $LaTiO_3$ films. As seen in Fig. 5c, the $Ti^{3+}$ spectra weight estimated from the fitting parameters of solid curves in Fig. 5b immediately reveal that in addition to the previously reported charge transfer from LTO into STO [49], there is an unexpected large charge transfer from YTO into STO which leads to a *localized electron layer at the YTO/STO interface* [50].

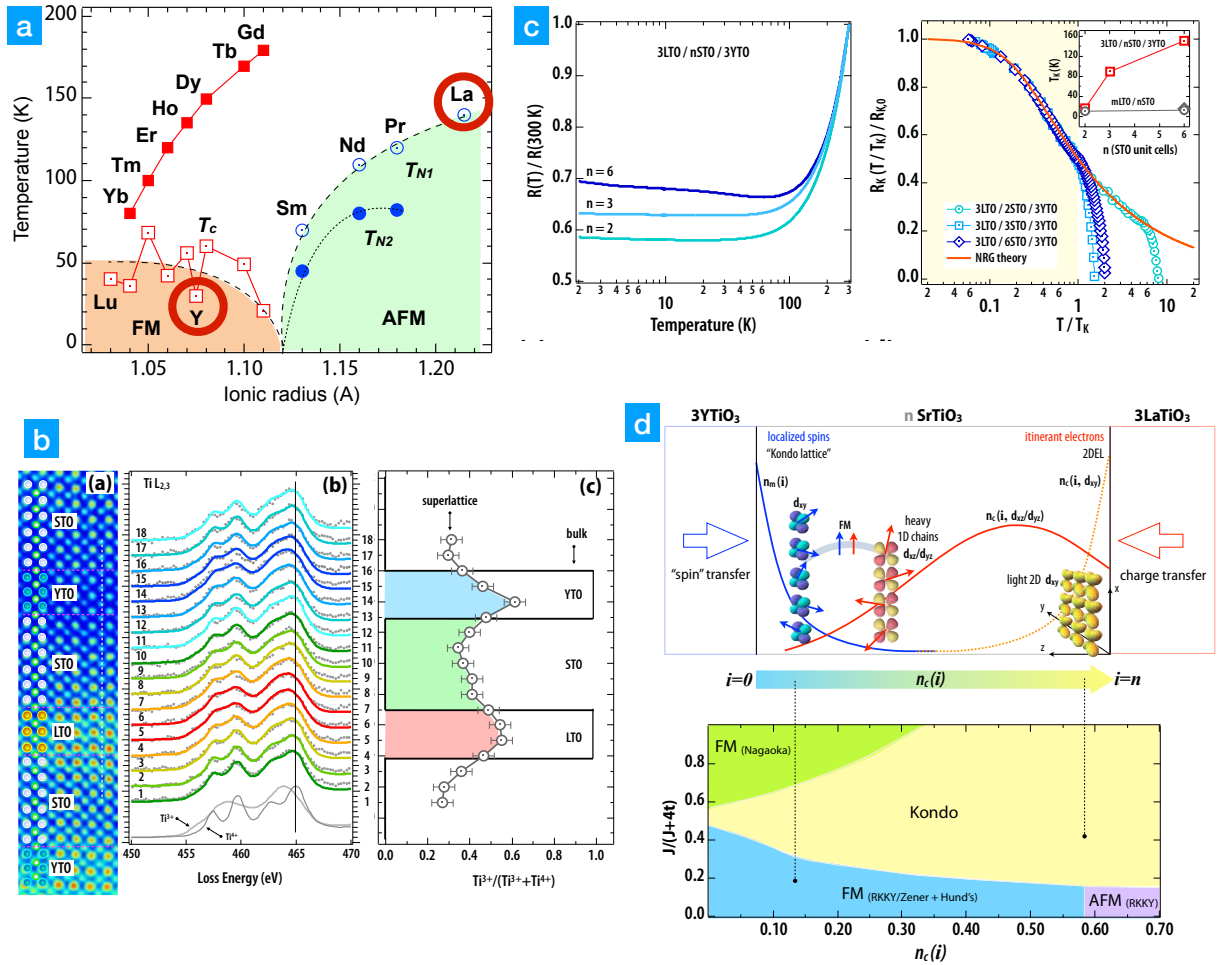

**Fig. 5:** *a: Magnetic phase diagram of $RTiO_3$. b: HAADF-STEM image of the tricolor superlattice reveals high-quality interfaces. STEM-EELS shows the layer-resolved chemical composition. c: All three samples display metallic behavior which agrees with theoretical fits. d: A view of the electron density distribution across the STO layer as dependent on layer thickness n and atomic plane i.*

Next, the authors investigated the properties of interfacial electrons arising from the interfacial charge transfer by measuring temperature-dependent electrical transport. As clearly seen in Fig. 5c all three samples 3LTO/$n$STO/3YTO (n=2,3,6) show characteristic metallic behavior with a weak upturn at a lower temperature. To rule out possible contributions from cation defects and oxygen vacancies, bilayer YTO/STO and LTO/STO samples were synthesized, and their transport properties were used as references [50, 51]. In sharp contrast to the highly insulating YTO/STO the sheet resistances of all the LTO/STO samples [51] show a *2D electron gas behavior*.

The other interesting feature seen in transport is the pronounced upturn in the sheet resistance at the lower temperature. Previous work on rare-earth titanate heterojunctions has attributed such an upturn to the Kondo lattice effect after carefully ruling out the contributions from weak localization [52] and electron-electron interactions [53]. One of the key features of the Kondo effect that immediately differentiates it from weak localization and electron-electron interactions is the *universal scaling behavior*. As shown in Fig. 5c, all tri-layer samples agree well with theoretical fits [solid red line, numerical renormalization group (NRG)] scaled Kondo resistances $[R_K(T/T_K)/R_{K,0}]$. The inset in Fig. 5c shows the extracted $n$-unit-cell dependent Kondo temperature T$_K$ by fitting the experimental data of [$m$LTO/$n$STO] and [3 u.c. LTO/$n$STO/3 u.c. YTO] superlattice. Based on the fact that the YTO/STO interface is highly insulating [50], the observed Kondo scaling behavior lends support to the formation of the interfacial lattice of localized magnetic moments located at the YTO/STO interface of LTO/STO/YTO. So overall, in the tri-color LTO/STO/YTO structures, the authors created metallic carriers at the LTO/STO interface facing the lattice of magnetic moments formed at the YTO/STO interface.

To better understand the magnetic interactions in the tri-color LTO/STO/YTO, we can look at the STO layer thickness ($n$-dependent) and atomic plane ($i$-dependent) electronic density distribution plus $d$-orbital occupancy across the STO layer. In other words the question is what happens when we move the 2DEG closer towards the magnetic lattice. The conceptual picture is given in the top panel of Fig. 5d.

First, because we deal with a two-dimensional electron gas, the carriers have very specific orbital types to let 'light' electrons rapidly move along the interface in the $x$-$y$-plane. In other words, for a thicker STO layer, we got $n_c(d_{xy}) \gg n_m$ near the LTO/STO interface, resulting in the formation of a Kondo singlet state (fully screened magnetic moments) since light $d_{xy}$ conduction electrons (dashed yellow line) with large carrier density are bound to the LTO/STO interface.

On the other hand, there is a low concentration of 'heavy' electrons (red arrows) with $d_{xz}$/$d_{yz}$ character (here $z$ is perpendicular to the interface), which slowly disperse away from the LTO/STO interface and appear near the magnetic STO/YTO interface (solid red line). Upon reaching the STO/YTO interface, these mobile heavy electrons interact with the localized magnetic moments $n_m(d_{xy})$. What is remarkable is that this *orbital-dependent ferromagnetic interaction* can proceed via two possible channels: (1) based on the Hund's rule, the interaction between the $d_{xy}$ and $d_{xz}$/$d_{yz}$ electrons results in the FM ground state [20] and (2) the Zener kinetic exchange, which can win the competition against the Kondo and RKKY interactions, again leading to the formation of a *localized ferromagnetic ground state with spin-polarized conduction electrons*. Finally, when the STO layer becomes ultra-thin, e.g., $n$=2, we have $n_c \sim n_m$ and the conduction carriers lose their distinct orbital character resulting in the mixed orbital state $d_{xz}/d_{yz}/d_{xy}$. In this case, in the ground state we have a direct competition between the Kondo screening, RKKY coupling, and Hund's energy. Based on this picture, the control of the STO thickness $n$ indeed modulates the critical ratio between magnetic sites and mobile carriers $n_m/n_c$ and their orbital character or orbital polarization to exert decisive control over the magnetic interactions.

## 5.3 New orbital order in graphene-like nickelates

In bulk, perovskite oxides have many exciting properties, including metal-insulator transitions, magnetism, superconductivity, charge and orbital orderings, and multi-ferroicity, to name a few. These infinite layer $ABO_3$ perovskite compounds consist of alternating $AO/BO_2$, $ABO/O_2$, and $AO_3/B$ atomic planes along the pseudo-cubic [001], [110], and [111] directions, respectively. Thus, the precise control during the growth of two or three pseudo-cubic (pc) unit cells of $ABO_3$ along the (111) direction leads to new lattice geometries with vertically shifted triangular planes of B sites and results in buckled honeycomb lattice as shown in Fig. 6a. The emergence of striking topological phases, including a quantum anomalous Hall state, was initially predicted for a honeycomb lattice by Haldane [54]. Recently, there was a spark of interest in the search for an *artificially stabilized graphene-like quasi-2D lattice* that can provide an ideal playground for interacting topological phases in complex oxides (for details, see [55]). Here I describe the case of rare-earth nickelates that illustrates the opportunities for designer topological phases by geometrical lattice engineering (GLE).

The first member of the rare earth nickelates series, $LaNiO_3$ (LN), is a paramagnetic metal. The other members of the family of nickelates ($RENiO_3$, $RE$=Pr, Nd,..., Lu, and Y) in bulk form exhibit metal-insulator transitions, E'-type antiferromagnetic ordering, charge ordering, and structural transitions with a strong dependence of the transition temperature on the size of the $RE$ ion. Several theoretical works (see Refs. [57–60] for review) further emphasized the possibility of realizing interaction-driven topological phases without spin-orbit ions (e.g., Dirac half semimetal phase, anomalous quantum Hall insulator phase, or ferromagnetic nematic phase) in the weakly correlated limit on the buckled honeycomb lattice of $RENiO_3$ as shown in Fig. 6b-c. Moreover, in sharp contrast to bulk $LaNiO_3$, where orbital ordering is absent, theoretical modeling in the strongly correlated limit predicted the presence of an *orbitally ordered magnetic phase as the novel ground state* for the buckled honeycomb lattice of $LaNiO_3$.

**How hard is it to grow (111) oriented films?** Despite the conceptual simplicity, the growth of perovskites along the [111]-direction presents a formidable challenge. Contrary to the conventional [001]-direction, the epitaxial stabilization along the [111] direction is far less understood due to unavoidable surface/chemical reconstruction effects. This can be seen as all perovskite substrates are strongly polar along this direction, e.g., $SrTiO_3$: $[SrO_3]^{4-}$, $Ti^{4+}$; $LaAlO_3$: $[LaO_3]^{3-}$, $Al^{3+}$, and so on. A possible solution to this polar catastrophe problem is to grow a thin metallic buffer layer at the beginning of the growth sequence to effectively screen the charge dipoles. However, one should pay particular attention as unwanted interfacial effects (between buffer layer and desired material) can significantly influence the buffered heterostructure.

As the reconstruction effect often appears only at the substrate and vacuum interfaces, choosing a substrate with the same sequence of charges per atomic plane as the desired material is another solution that does not require growing a metallic buffer layer. To investigate this, Middey *et al.* have grown $LaNiO_3$ films on (111) $SrTiO_3$ (with a polar jump at the film/substrate interface) and compared it to the growth of $LaNiO_3$ on (111) oriented $LaAlO_3$ (without any polar jump at the film/substrate interface) [56]. It was clearly demonstrated that while a thick bulk-like $LaNiO_3$ film is metallic and effectively screens charge dipoles, a thinner film becomes insulating. Using X-ray diffraction (XRD) and X-ray absorption spectroscopy (XAS), the authors uncovered massive amounts of oxygen vacancies within the thinner films. With the increased LNO thickness, the increased metallicity screens the polar jump, and the relative amount of proper $Ni^{3+}$ ions rapidly increases. Finally, good stoichiometric $LaNiO_3$ along [111] can be only obtained when the film thickness reaches 15 unit cells. In sharp contrast, $Ni^{3+}$ was stabilized from the very initial stage of growth of LNO on the LAO (111) substrate and thus confirmed that the *absence of a polar jump at the film-substrate interface* is critically important for the epitaxial stabilization along [111]. As a result, the desired generalized *graphene-like crystal of $RENiO_3$* with *RE*=La to Nd has been successfully achieved for the first time on the LAO (111) substrate [60].

**What about the ground state?** According to theoretical calculations, the QAH topological phase should be accompanied by spontaneous ferromagnetism. However, X-ray resonant magnetic scattering measurements on (111) oriented [2u.c. $NdNiO_3$/4u.c. $LaAlO_3$] (2NNO/4LAO after that) superlattice ruled out the possibility of a long-range ferromagnetic ground state and instead established the presence of *antiferromagnetic* correlations (see Fig. 6d). In addition to magnetism, the orbital structure was investigated by the X-ray linear dichroism (XLD) technique. The XLD spectroscopy allows for uncovering different kinds of orbital ordering and the symmetry of a specific orbital state per chemical element of the film. In this resonant X-ray spectroscopy, one measures the difference in absorption with vertical polarization $I_H$ vs. horizontal one $I_V$. If orbitals are preferentially aligned along one of the X-ray polarizations, the XLD signal becomes strongly enhanced. Conversely, for the orbitally disordered state, the X-ray linear dichroism is zero.

Despite its conceptual simplicity, the geometrical arrangement between the sample and X-ray polarization vector requires careful consideration for detecting orbital ordering. Specifically, as illustrated in Fig. 6e, all the $e_g$ orbitals ($3d_{z2-r2}$, $3d_{x2-r2}$, and $3d_{y2-r2}$) are oriented at $\phi = 54.7^o$ with respect to the [111] growth axis of a $NiO_6$ octahedron. Because of this specific geometrical arrangement, the XLD signal is expected to be very small even for a ferro-orbital ordered (FOO) state with 100% orbital polarization.

To maximize the XLD signal, the samples can be mounted on a copper wedge (Fig. 6f), which 'mechanically' reorients the Ni-O bonds along the vertical polarization V and the in-plane horizontal polarization H. This specific sample orientation on the wedge aligns the $3d_{z2-r2}$ orbitals almost along V polarization, giving a finite dichroic signal. On the other hand, $3d_{x2-r2}$ orbitals are almost in the plane of H polarization with a small but finite angle with respect to the polarization vector H, resulting in an opposite and strongly reduced XLD signal. As a result, instead of perfect cancellation of linear dichroism, a finite XLD is expected to be present for the antiferro-orbital ordered (AFO) state. Fig. 6e-f shows resonant Ni $L$-edge X-ray absorption spectra (XAS) ($I_V$ and $I_H$) and the XLD spectra (difference between $I_V$ and $I_H$) obtained in the flat ($\phi = 0^o$) and the wedge ($\phi = 45^o$) geometries.

As anticipated from the discussion above, the XLD signal indeed strongly increased when the measurement was conducted for the $\phi = 45^o$ geometry. As a control experiment, for the thick bulk-like (111) NNO film, the XLD measurement shows no significant orbital polarization.

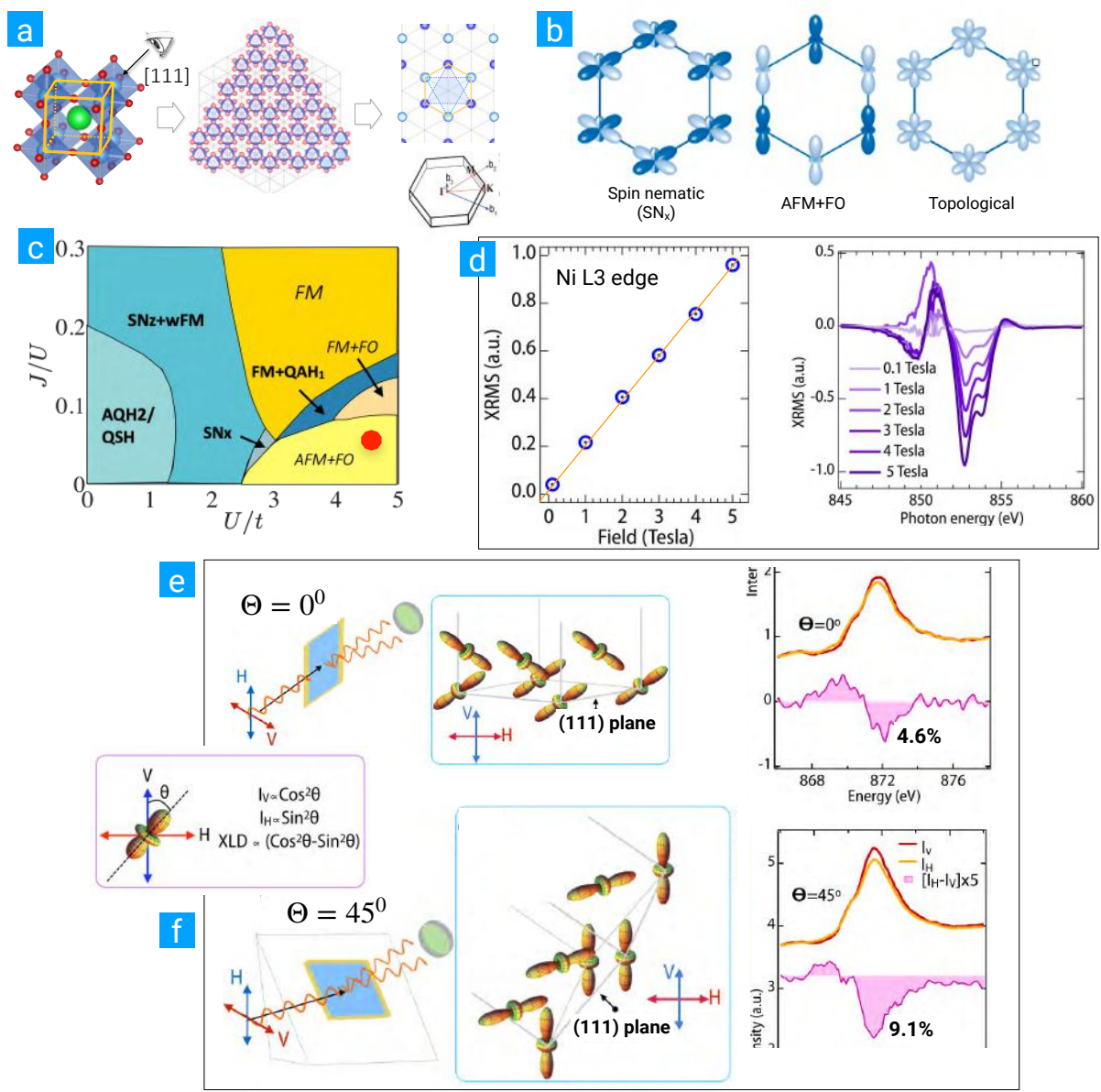

**Fig. 6:** *a: Growth of $ABO_3$ along [111] leads to new lattice geometries. b-c: The weakly-correlated limit of $RENiO_3$ has been theorized to host interaction-driven topological phases such as a ferromagnetic nematic phase, Dirac half semimetal, and anomalous quantum Hall insulator. d: XRMS measurements on 2NNO/4LAO establish the presence of antiferromagnetic correlations. e-f: XLD spectra for $\phi = 0°$ (flat) and $\phi = 45°$ (wedge) geometries. The XLD signal was greatly increased for the wedge geometry, a sign of orbital polarization in 2NNO/4LAO.*

This control experiment emphasizes that the observed orbital polarization in the 2NNO/4LAO (111) superlattice is not a measurement artifact, and this buckled honeycomb lattice geometry engineered the orbitally polarized ground state. The obtained value of XLD around 9% is large as the finite bandwidth of the $e_g$ bands and strong covalency strongly reduce the orbital polarization from the atomic limit. I need to mention, however, that by the nature of the spectroscopic probe, XLD can only establish the presence of orbital ordering or orbital polarization but cannot resolve a specific type of orbital pattern present in the system. This can be done by using synchrotron-based resonant X-ray scattering on the Ni $L$-edge and by using density functional theory (DFT). Combined with the DFT prediction the experimental data revealed a novel kind of *anti-ferro-orbital ordering with staggered* $3d_{z^2-r^2}$ *orbitals rotated by 90$^o$ in subsequent layers*. This new quantum state is absent in the bulk nickelates.

## 5.4 Electronic structure of graphene-like nickelates

*Probing buried graphene-like [111] perovskite layers using soft and hard X-ray photoemission.* As we have seen in the previous section, artificial heterostructures comprised of ultra-thin complex-oxide layers grown in the pseudo-cubic [111] direction have been predicted to harbor a wide range of extraordinary quantum states stemming from the unique lattice geometry resembling graphene and the interplay between strong electronic correlations and band topology [60]. However, studying such atomic layers' electronic and structural properties remains a formidable challenge due to the limitations of conventional surface-sensitive techniques, which typically probe depths of only a few Angstroms.

In this section, I want to discuss a new experimental methodology that combines bulk-sensitive soft X-ray angle-resolved photoelectron spectroscopy (SX-ARPES), hard X-ray photoelectron spectroscopy (HAXPES), and first-principles DFT+U calculations. This powerful set is used to establish a direct and reliable method for *extracting momentum-resolved and angle-integrated valence-band electronic structures* of an ultra-thin buckled graphene-like layer of $NdNiO_3$-(111) sandwiched between two 4-unit-cell thick layers of insulating $LaAlO_3$-(111) (see Fig. 6a). Clearly, this is a challenging system for measurements as the active monolayer of $NdNiO_3$-(111) is buried under the capping layer of $LaAlO_3$-(111) [61].

First, soft X-ray angle-resolved photoemission spectroscopy (SX-ARPES) measurements shown in Figure 7b were performed using the high-resolution ADRESS beamline at the Swiss Light Source. To enhance the information depth and enable a momentum-resolved analysis of the hexagonal $NdNiO_3$, the measurements were conducted at high incoming photon energies ranging from 642 to 874 eV, far beyond conventional UV ARPES. Using such high photon energies effectively increases the inelastic mean-free paths of the photoelectrons within the superlattice by a factor of 3–5 compared to typical home ARPES investigations, substantially improving the probing depth. Theoretical calculations suggest the existence of two possible orbital arrangements in this material: one characterized by P1 symmetry in a $1\times1$ unit cell (P1 $1\times1$) and another featuring P3 symmetry in a larger $\sqrt{3}\times\sqrt{3}R30^o$ lateral unit cell. Bulk-sensitive momentum-resolved measurements reveal excellent agreement with the band structure calculated using DFT+U for the Ni sites' antiferromagnetic (AFM) ordering with P1 $1\times1$ symmetry shown in Figure 7c. These measurements provide direct evidence supporting the P1 $1\times1$ symmetry, which perfectly aligns with the findings previously suggested by X-ray linear dichroism data (see Sec. 5.3) [61–63].

To investigate the entire sample depth, angle-integrated hard X-ray photoemission spectroscopy (HAXPES) measurements of the valence bands were carried out at a photon energy of 6.45 keV.

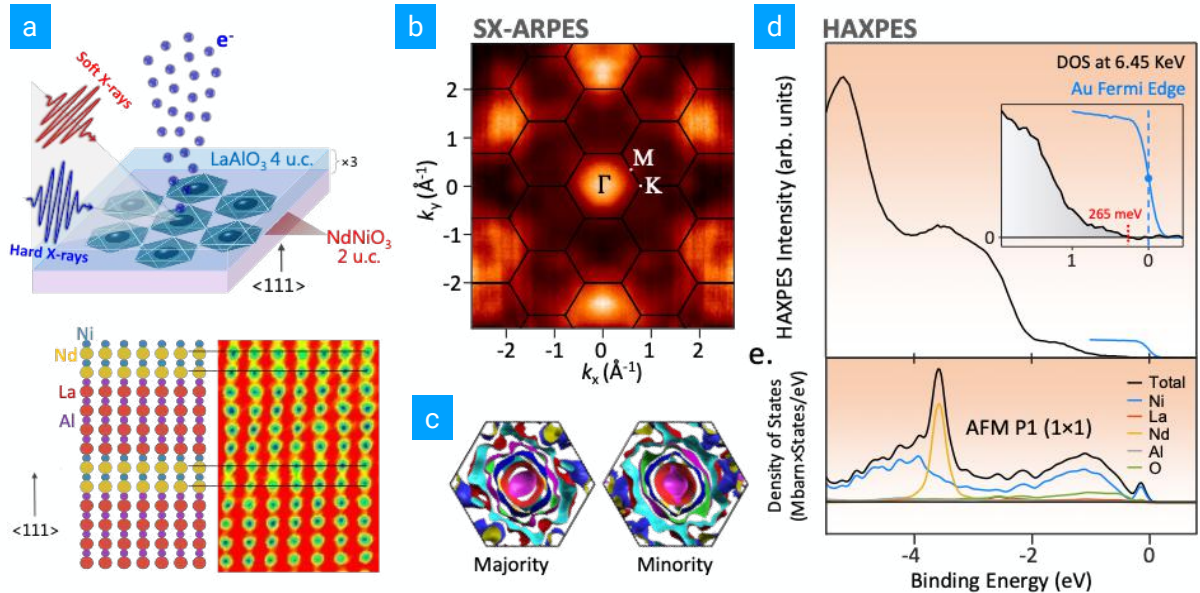

**Fig. 7:** *a: Schematic diagram of the sample and the SX-ARPES and HAXPES experimental geometries. b: Momentum-resolved SX-ARPES map of the Ni 3d states near the Fermi level. c: Isoenergetic cuts through the band structure in reciprocal space for the majority and minority bands. d: Bulk-sensitive HAXPES spectrum of the valence-band DOS. Inset shows a high-statistics spectrum of the valence-band maximum (at ∼ 265 meV), referenced to the Au Fermi edge. e: Cross-section-weighted element-projected and total DOS of the superlattice, calculated in the GGA+U framework of DFT and broadened to account for both experimental and hole lifetime broadening.*

At this hard X-ray energy, the estimated inelastic mean-free path is approximately 9 nanometers, enabling direct probing of the density of states across the entire sample. This approach facilitates a straightforward comparison between experimental data (Fig. 7d) and theoretical predictions (Fig. 7e). The experimental HAXPES valence-band spectrum shows remarkable agreement with the first-principles calculations regarding relative intensities and the positions of key features. Moreover, the experimental valence-band maximum is observed at a binding energy of ∼265 meV below the Fermi level. This value corresponds to the magnitude of the valence-band bandgap, indicating that the full bandgap of $NdNiO_3$-(111) is at least of this size. I remind the reader that in bulk (above 150 K), $NdNiO_3$ is a paramagnetic metal [61–63].

In conclusion, a combination of bulk-sensitive soft and hard X-ray photoemission techniques can be utilized to investigate the *momentum-resolved electronic band dispersion of a buried* two-dimensional $NdNiO_3$-(111) layer within a designed superlattice below a cap layer. Additionally, the density of states of this structure can be directly measured using HAXPES. Combined with first-principles DFT+U calculations, this new methodology provides direct and definitive evidence for an antiferro-orbital (AFO) order characterized by P1 symmetry within a 1×1 unit cell [64].

We learn from this example that we finally have a practical experimental approach to investigate the *momentum resolved* electronic structure of new quantum metals and semimetals as thin as a single atomic plane.

## 5.5　Artificial quantum spin liquid on lattices with extreme frustration

**Can we make a new exotic state of matter?** In what follows, I will discuss the, in my view, most enigmatic yet least experimentally understood state of matter called a spin liquid. To make things even more intriguing, we should look at its quantum version, a quantum spin liquid or QSL. If there is something that we do not fully understand microscopically, it would be the concept of liquid (e.g., water). These days, experimentalists are very good at finding and describing long-range order (LRO) by sharp Bragg intensities in reciprocal space. Conversely, a disordered spin gas phase (paramagnet) can be reliably detected. But the precise nanoscopic description of something which fluctuates in space and time with numerous short-range ordered (SRO) configurations of spins is still beyond our current computational capabilities.

Nevertheless, one can bravely embrace the idea and think, at least theoretically, if such a liquid of quantum spins is thermodynamically stable, what would be a Landau-kind order parameter if at all, and most importantly, how could we detect such a state experimentally (for a theory discussion see [65–70]).

As for the question of stability, we have one example of a true quantum spin liquid, $^3$He, which exists only outside of the solid-state setting. Even theoretically, we still do not have complete answers to those questions. For example, we do know that a QSL cannot be described by the conventional Landau theory of phase transitions relying on spontaneously broken symmetry. Instead, one can introduce the idea of 'entanglement entropy' as a topological order parameter. Despite its novelty and usefulness for theory, experimentally, we do not have probes that couple to such a 'topological order parameter.' Thus, we mostly rely on negative statements about a quantum material in question for practical reasons. At best, we can verify in a magnetometer or via some sort of magnetic scattering that our magnetic crystal has no long-range spin order down to the lowest experimentally accessible temperature. This is hardly a satisfactory situation, but we need to say what would constitute a set of positive statements as an alternative. Here I list a few popular ones:

- Definition 1: a QSL is a state in which the spin-spin correlations decay to zero at large distances or $\langle S_i^\alpha S_j^\beta \rangle \to 0$ when $|r_i - r_j| \to \infty$. Objection: A classical liquid, spin nematic, or valence bond crystal also satisfies this definition.

- Definition 2: a QSL is a state without any spontaneously broken symmetry, but so is a valence-bond solid.

- Definition 3: a QSL is *a Mott insulator that possesses no long-range magnetic order, lacks any spontaneously broken symmetry, and carries a spectrum of fractional excitations*. At present, this is the definition most amicable for experimental verification. I recap that fractional excitations are quasi-particles, e.g., spinons, carrying a half-odd-integer spin, and fractionalized fermions are coupled to an emergent gauge field.

**Why do we care about such exotics?** Here is a short list of reasons: (1) Most SQLs are 'flat-band' systems; if doped, they may harbor high or potentially even room-temperature superconductivity, (2) for dimensionality greater than one, fractional excitations interact with each other through emerging gauge fields, giving rise to string- and loop-like excitations akin to physics of quark-gluon plasma and (3) QSLs sustain a new type of topological non-local order and new spin excitations (anyons) which can be useful as an unconventional platform for quantum computing with topological qubits beyond silicon, aluminum, or ion traps.

**How to make a QSL?** As for the experimental realization of a QSL, the currently existing "recipes" are illuminating but very limited. On the one hand, a general guiding principle is that to reach a QSL, significant frustration resulting from the lattice geometry, multiple exchange terms, or bond conflict are the most essential prerequisites. After tremendous, decades-long efforts, promising *candidate materials* have been proposed and synthesized [71]. Interestingly, the underlying lattices of almost all known QSL candidates are bound to five types of geometries: triangular, pyrochlore, kagome, hyperkagome, and honeycomb lattices. This, in turn, limits the pursuit of new QSL materials and brings to the focus an open question of whether any additional lattice motifs can host a QSL and how it can be achieved experimentally?

Figure 8a illustrates the new approach for making such exotic phases called *geometrical lattice engineering* (GLE) (see Sec. 3 and Ref. [22]). GLE principally aims to design and fabricate lattices of artificial geometry by stacking on demand a specific number of atomic planes along unconventional crystallographic directions to facilitate unattainable in the bulk configuration of charges, orbitals, and spins. You had already seen this concept in action in subsection 5.3 where I described how to create graphene-like $NdNiO_3$ with a new anti-ferro orbital order.

**Is there a real QSL based on the GLE?** Here, I introduce a generic design of a new (quasi-2D) lattice derived from the spinel structure (chemical formula $AB_2O_4$) and demonstrate its feasibility for a QSL phase [72]. Concretely, I will use $CoCr_2O_4$ as a prototype; we fabricated a series of (111)-oriented ultra-thin films confined by non-magnetic $Al_2O_3$ layers into a quantum well geometry. Compared to its bulk counterpart, the onset of the ferrimagnetic transition decreases monotonically with reduced thickness and eventually shuts off in a single-unit slab of (111) $CoCr_2O_4$. In this quasi-2D limit, the degree of magnetic frustration becomes enhanced by almost 3 orders of magnitude with persisting spin fluctuations down to 30 mK. $CoCr_2O_4$ belongs to the normal spinel chromite family, $MCr_2O_4$ ($M$=Mn, Fe, Co, and Ni) where the magnetically active $M^{2+}$ ions occupy the tetrahedral A sites of the diamond sublattice and the $Cr^{3+}$ ions occupy the octahedral B sites of the pyrochlore sublattice, possessing complex spin configuration of the ground state. Note, in bulk, $CoCr_2O_4$ has a collinear ferrimagnetic state first formed with the Curie temperature of ∼93 K, it transforms into an incommensurate spiral ferrimagnetic state at ∼26 K, and finally, an incommensurate to commensurate lock-in transition takes place at ∼14 K.

Now on to the GLE. As seen in Fig. 8a/b, when viewed along the [111] direction, the structure is an intrinsic stacking of triangle (T) and Kagome (K) cation planes from Co and Cr ions embedded in the oxygen cubic close-packed frame. This leads to a sequence '-O-Cr(K)-O-Co(T)-Cr(T')-Co(T)-' in a single unit with four cation layers, which we denote as one quadruplet layer (1 QL). Based on this design idea, [$n$ QL $CoCr_2O_4$/1.3 nm $Al_2O_3$] ($n$=1, 2, 4; 1 QL ∼ 0.48 nm) superlattices were fabricated by pulsed laser deposition on (0001)-oriented single crystal $Al_2O_3$ substrates.

**What is the ground state of this new synthetic magnet?** The investigation of the magnetic behavior of each sublattice was done by recording the resonant X-ray absorption spectroscopy (XAS) taken with left- and right-circularly polarized beams. The difference between those two spectra, called X-ray magnetic circular dichroism (XMCD), reflects the net magnetization of a specifically probed element. To make the statement even more contrasty, I will l compare the 4QL sample (bulk-like) to the most intriguing 1QL sample. As seen in Fig. 8c, the saturated magnetic signal is indeed observed in the thicker 4 QL $CoCr_2O_4$ and also it exhibits clear hysteresis loops at both Co and Cr $L_3$ edges. These findings signal for the long-range magnetic order even in 4QL thin samples. However, in sharp contrast, no hysteresis loop but a linear XMCD versus H relationship is found on both Co and Cr in 1 QL $CoCr_2O_4$, typical of a paramagnetic behavior!

To further examine if any long-range spin ordering emerges at extremely low temperatures, we performed the torque magnetometry measurements on 1 QL $CoCr_2O_4$ from 30 K down to 30 mK. This technique quantifies the magnetic torque response of a sample with respect to the applied magnetic field ($\tau \propto M \times H$). It is an exquisitely sensitive utility to probe vanishingly small magnetic signals from ultra-thin samples and interfaces. The result confirms that within the resolution of measurement and the entire temperature range, no hysteresis but a reversible parabolic $\tau \propto (\mu_0 H)^2$ relationship is observed for 1 QL, which implies a quantum paramagnetic behavior persisting down to 30 mK. Moreover, if we plot torque vs. $T$ (see Fig. 8c (right)), we can immediately see that even at 30 mK, the ground state has large spin entropy, and its spectrum of excitations is *gapless* or spin-metal like. To make this story even more compelling, an extensive set of neutron reflectivity and mount spin resonance data affirms these conclusions.

**Why does this quasi-2D artificial magnet enter a gapless QSL state?** To obtain a microscopic insight into *how the designer lattice topology and quantum confinement alter the exchange interactions* and, consequently, the magnetic ground state, a set of DFT calculations and Monte-Carlo simulations were performed on bulk, 2, and 1 QL of (111) $CoCr_2O_4$. As seen in Fig. 8d (left), the theory revealed that the new emergent QSL ground state is a consequence of the markedly smaller exchange interaction along (111) because it is blocked by the vacuum-like spacer of $Al_2O_3$ with a gap of 4 eV in the ultrathin films compared to bulk. In fact, for 1 QL, this interaction is completely suppressed (!), while the in-plane interactions remain essentially bulk-like, all in contrast to the behavior in bulk. And now we understand that due to the entire blocked exchange along (111), our 1QL magnet turned into a 2D system with a pure Kagome-triangular motive of extremely large magnetic frustration. Fig. 8d (right) shows that the system reaches the frustration factor $f = T_{CW}/T_N$ of about 1000 (!) compared to the value in the bulk of 4–6.

Here is what we learned from GLE. The ground state is a well-defined long-range magnetically ordered state in bulk (QL $\to \infty$). As the number of QLs is reduced, it becomes more and more difficult to stabilize a conventional ordered state due to the enhancement of magnetic frustration. Eventually, the ground state becomes highly degenerate in 1 QL, unleashing dynamical spin fluctuations. I remind you that this is the regime where quantum effects play a pivotal role in bringing the system into a QSL (quantum paramagnet) state without a spin gap.

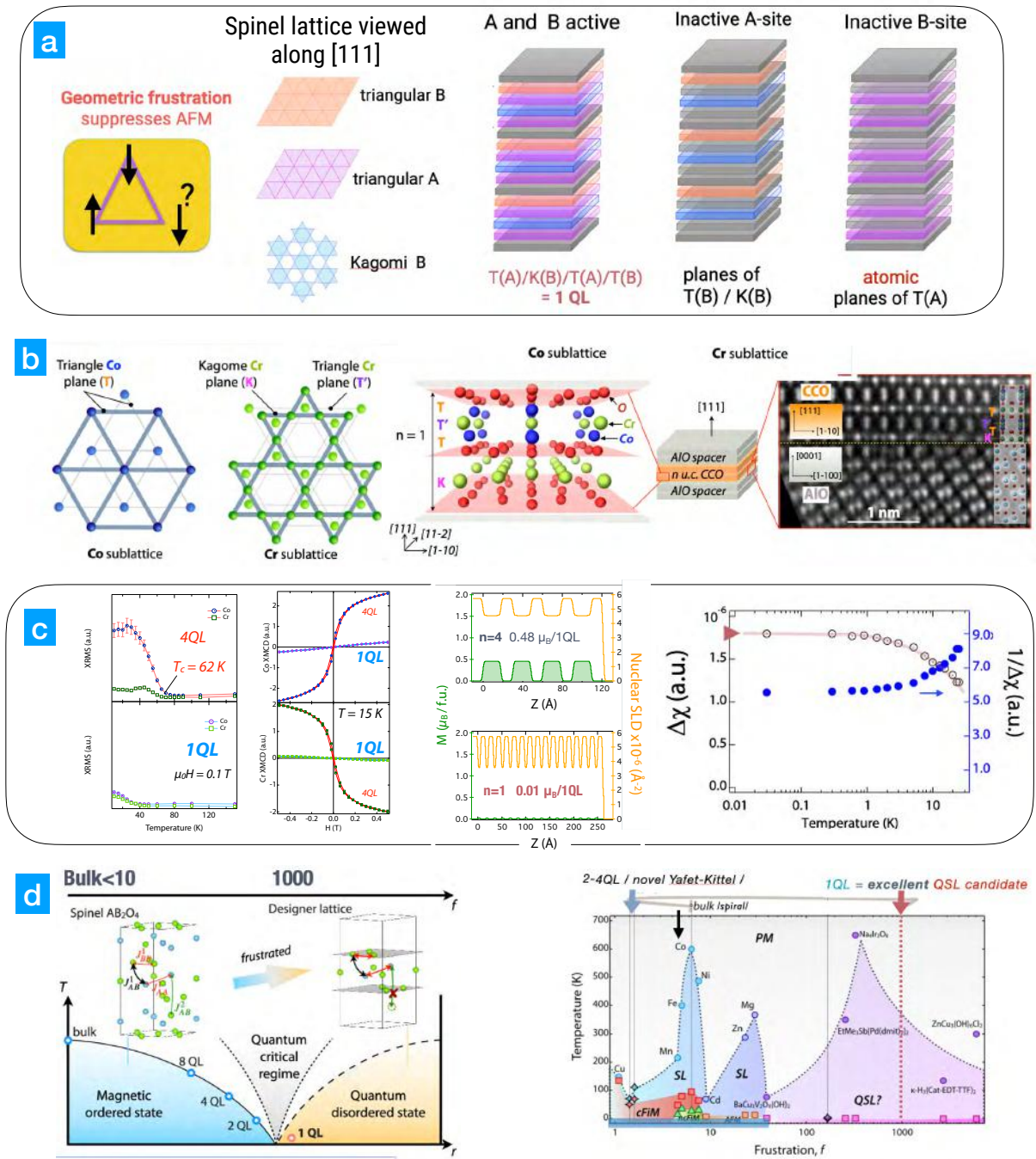

**Fig. 8:** *a: The spinel lattice is made of a stack of triangle and kagome planes when viewed along the [111] direction. b: View of cation triangle and kagome planes. c: XMCD measurements show long-range magnetic ordering even in 4QL CCO films, but paramagnetic behavior in 1QL CCO. Torque measurements reveal that even at 30 mK, the 1QL sample is gapless. d: Theory reveals the emergent QSL ground state is a consequence of the suppression of out-of-plane exchange interaction.*

# 6   Problems to solve, ideas to try

In this section, I want to point in ten directions, which is my challenge to you. Frankly, I do not have answers to any of those questions, so I leave it to you to seek possible solutions.

1. If you replace a conventional nano-second UV laser with a femtosecond one, what new synthesis regime can you reach? The intrigue here is that in the femtosecond regime, there is no time for laser heat dissipation as phonons are too slow (pico-second timescale).

2. How can we grow uniaxially strained structures?

3. Apart from GLE, how do we design structures with extreme frustration from interactions?

4. Can you think of a way to combine different classes of interesting quantum materials, such as TMO with TM dichalcogenides, TM oxyfluoride, and TM nitrites?

5. What happens if you combine different topological classes and antagonistic orders such as Dirac electrons-Cooper pairs or Cooper pairs and magnetic monopoles of a spin-ice?

6. Can you think of a design approach for structures that can 'zoom in' on a specific term of a Hamiltonian?

7. Can you create structures holding quantum chaos?

8. What about structures that reach a regime of quantum hydrodynamics?

9. How can you measure a spectrum of excitations (e.g., magnons, phonons, plasmons, orbitons) right at the interface?

10. What designer structures can directly reveal the quantum entanglement of fermions?

## Acknowledgments

I dedicate this lecture to Prof. Daniel Khomskii, for whom I am indebted for many ideas described throughout the text and who was and still is a great source of influence and inspiration for what it means to be a physicist. Also, I express my deep gratitude to my postdoctoral advisor Prof. Bernhard Keimer who relentlessly shared his encyclopedic knowledge of condensed matter physics and for his guidance in my first steps into the physics of oxide interfaces. And finally, I offer my sincere thanks to all my Ph.D. students, postdocs, and collaborators. They contributed countless research hours and incredible talents in pushing the frontiers of this stimulating field. The introductory section 1 is derived in part from Ref. [73]. Work on this manuscript was supported by the U.S. Department of Energy, Office of Science, Office of Basic Energy Sciences under Award Number DE-SC0022160.

Oxide Interfaces 9.29

# References

[1] B. Keimer and J.E. Moore, Nat. Phys. **13** 1045 (2017)

[2] D.I. Khomskii: *Transition metal compounds* (Cambridge University Press, 2014)

[3] L. Pauling, J. Am. Chem. Soc. **51** 1010 (1929)

[4] L. Pauling: *The nature of the chemical bond and the structure of molecules and crystals: an introduction to modern structural chemistry* (Cornell University Press, 1960)

[5] P. Fazekas: *Lecture Notes on Electron Correlation and Magnetism* (World Scientific, Singapore, 1999)

[6] M.C. Gutzwiller, Phys. Rev. Lett. **10** 159 (1963)

[7] J. Kanamori, Prog. Theor. Phys. **30** 275 (1963)

[8] J. Hubbard, Proc. Roy. Soc. London A **277** 237 (1964)

[9] D.I. Khomskii and S.V. Streltsov, Chem. Rev. **121** 2992 (2020)

[10] D.I. Khomskii: *Basic aspects of the quantum theory of solids: order and elementary excitations* (Cambridge University Press, 2010)

[11] E. Pavarini, E. Koch, J. van den Brink, and G. Sawatzky (eds.): *Quantum Materials: Experiments and Theory*, Modeling and Simulation, Vol. 6 (Forschungszentrum Jülich, 2016)

[12] W. Witczak-Krempa *et al.*, Annu. Rev. Condens. Matter Phys. **5** 57 (2014)

[13] J. Chakhalian *et al.*, Rev. Mod. Phys. **86** 1189 (2014)

[14] H.Y. Hwang *et al.*, Nat. Mater. **11** 103 (2012)

[15] P. Zubko *et al.*, Annu. Rev. Condens. Matter Phys. **2** 141 (2011)

[16] S. Stemmer and S.J. Allen, Annu. Rev. Mater. Res. **44** 151 (2014)

[17] K.R. Poeppelmeier and J.M. Rondinelli, Nat. Chem. **8** 292 (2016):

[18] J. Chakhalian, A.J. Millis, and J. Rondinelli, Nat. Mater. **11** 92 (2012)

[19] J. Liu *et al.*, Nat. Commun. **4** 2714 (2013)

[20] J.W. Freeland *et al.*, Europhys. Lett. **96** 57004 (2011)

[21] J. Chakhalian *et al.*, Nat. Phys. **2** 244 (2006)

[22] X. Liu *et al.*, MRS Commun. **6** 133 (2016)

[23] A.M. Cook and A. Paramekanti, Phys. Rev. Lett. **113** 077203 (2014)

[24] M. Kareev *et al.*, J. Appl. Phys. **109** 114303 (2011)

[25] G. Koster *et al.*: *Growth studies of heteroepitaxial oxide thin films using reflection high-energy electron diffraction (RHEED)* in G. Koster, M. Huijben, and G. Rijnders (eds.): *Epitaxial Growth of Complex Metal Oxides* (Woodhead Publishing, 2015) pp.3–29

[26] G. Koster: *Reflection high-energy electron diffraction (RHEED) for in situ characterization of thin film growth* in G. Koster and G. Rijnders (eds.): *In situ characterization of thin film growth* (Woodhead Publishing, 2011) pp.3–28

[27] P.W. Anderson and E.I. Blount, Phys. Rev. Lett. **14** 217 (1965)

[28] N.A. Benedek and T. Birol, J. Mater. Chem. C **4** 4000 (2016)

[29] Y. Shi *et al.*, Nat. Mater. **12** 1024 (2013)

[30] T.H. Kim *et al.*, Nature **533** 68 (2016)

[31] T. Kolodiazhnyi *et al.*, Phys. Rev. Lett. **104** 147602 (2010)

[32] J. Fujioka *et al.*, Sci. Rep. **5** 13207 (2015)

[33] T. Ishidate *et al.*, Phys. Rev. Lett. **78** 2397 (1997)

[34] M. Mochizuki and M. Imada, New J. Phys. **6** 154 (2004)

[35] Y. Cao *et al.*, Phys. Rev. Lett. **116** 076802 (2016)

[36] A. Ohtomo *et al.*, Nature **419** 378 (2002)

[37] Y.J. Chang *et al.*, Phys. Rev. Lett. **111** 126401 (2013)

[38] J. Kondo, Prog. Theor. Phys. **32** 37 (1964)

[39] A.C. Hewson: *The Kondo problem to heavy fermions* (Cambridge University Press, 1997)

[40] G.-Y. Guo, S. Maekawa, and N. Nagaosa, Phys. Rev. Lett. **102** 036401 (2009)

[41] H. Tsunetsugu, M. Sigrist, and K. Ueda, Rev. Mod. Phys. **69** 809 (1997)

[42] S. Doniach, physica B+C **91** 231 (1977)

[43] T. Jungwirth *et al.*, Rev. Mod. Phys. **78** 809 (2006)

[44] P. Fazekas and E. Müller-Hartmann, Z. Phys. B **85** 285 (1991)

[45] C. Zener, Phys. Rev. **81** 440 (1951)

[46] J. Mannhart and D.G. Schlom, Science **327** 1607 (2010)

[47] T. Katsufuji, Y. Taguchi, and Y. Tokura, Phys. Rev. B **56** 10145 (1997)

[48] Y. Cao *et al.*, Appl. Phys. Lett. **107** 112401 (2015)

[49] H.W. Jang *et al.*, Science **331** 886 (2011)

[50] Y. Cao *et al.*, Nat. Commun. **7** 10418 (2016)

[51] See Supplemental Material of [35] for details and additional data.

[52] S.-P. Chiu and J.-J. Lin, Phys. Rev. B **87** 035122 (2013)

[53] M. Gabay and J.-M. Triscone, Nat. Phys. **9** 610 (2013)

[54] D.N. Sheng *et al.*, Phys. Rev. Lett. **97** 036808 (2006)

[55] X. Liu *et al.*, MRS Commun. **6** 133 (2016)

[56] S. Middey *et al.*, Sci. Rep. **4** 6819 (2014)

[57] D. Xiao *et al.*, Nat. Commun. **2** 596 (2011)

[58] K.-Y. Yang *et al.*, Phys. Rev. B **84** 201104 (2011)

[59] D. Doennig, W.E. Pickett, and R. Pentcheva, Phys. Rev. B **89** 121110 (2014)

[60] G.A. Fiete and A. Rüegg, J. Appl. Phys. **117** 172602 (2015)

[61] A. Arab *et al.*, Nano Lett. **19** 8311 (2019)

[62] L. Plucinski *et al.*, Phys. Rev. B **78** 035108 (2008)

[63] S. Middey *et al.*, Phys. Rev. Lett. **116** 056801 (2016)

[64] A.X. Gray *et al.*, Nat. Mater. **10** 759 (2011)

[65] L. Savary and L. Balents, Rep. Prog. Phys. **80** 016502 (2016)

[66] J. Knolle and R. Moessner, Annu. Rev. Condens. Matter Phys. **10** 451 (2019)

[67] Y. Zhou, K. Kanoda, and T.-K. Ng, Rev. Mod. Phys. **89** 025003 (2017)

[68] S. Sachdev, Nat. Phys. **4** 173 (2008)

[69] L. Balents, Nature **464** 199 (2010)

[70] F. Mila, Eur. J. Phys. **21** 499 (2000)

[71] J.R. Chamorro, T.M. McQueen, and T.T. Tran, Chem. Rev. **121** 2898 (2020)

[72] X. Liu *et al.*, Nano Lett. **21** 2010 (2021): 2010

[73] X. Liu: *Artificial Quantum Many-Body States in Complex Oxide Heterostructures at Two-Dimensional Limit* (PhD Thesis, University of Arkansas, 2016)